# THE PROBLEM OF MEASUREMENT AND THE THEORY OF QUANTUM STATE REDUCTION

Masanao Ozawa

*School of Informatics and Sciences, Nagoya University, Chikusa-ku, Nagoya 464-01, Japan*

**Abstract**

A new approach to the problem of measurement in quantum mechanics is proposed. In this approach, the process of measurement is described in the Heisenberg picture and divided into two stages. The first stage is to transduce the measured observable to the probe observable. The second stage is to amplify the probe observable to the macroscopic meter observable. Quantum state reduction is derived, based on the quantum Bayes principle, from the object-apparatus interaction in the first stage. The dynamical process of the second stage is described as a quantum amplification with *infinite gain* based on nonstandard analysis.

## 1 Introduction

A model of quantum measurement is specified by how the apparatus is prepared, how it interacts with the object, and how the outcome is obtained from it. On the other hand, a quantum measurement is specified from a statistical point of view by the outcome probability distribution and the state reduction, the state change from the state before measurement to the state after measurement conditional upon the outcome. If two measurements have the same outcome probability distribution and the same state reduction, they are statistically equivalent.

The conventional derivation of the state reduction from a given model of measurement is to compute the state of the object-apparatus composite system just after the measuring interaction and to apply the projection postulate to the subsequent measurement of the probe observable. The validity of this derivation is, however, limited or even questionable, apart from the interpretational questions such as "von Neumann's chain" argument [1], because of the following reasons:

1. The probe detection, such as photon counting, in some measuring apparatus does not satisfy the projection postulate [2].

2. When the probe observable has continuous spectrum, the projection postulate to be applied cannot be formulated properly in the standard formulation of quantum mechanics [3, 4].

3. When another measurement on the same object follows immediately after the first measurement, the measuring apparatus for the second measurement can interact with the object just after the measuring interaction, and just before the probe measurement, for the first measurement. The state reduction obtained by the conventional approach determines the state just after the probe measurement for the first measurement and cannot give the joint probability distribution of the outcomes of the above consecutive measurements [5].

A mathematically rigorous and physically consistent derivation of the state reduction from any model of measurement without appealing to the projection postulate to the probe measurement has been established in [3, 4, 6]. Based on this derivation, the statistical equivalence classes of all the possible quantum measurements are characterized as the normalized completely positive map valued measures [3, 6].

In this paper, a new approach to the problem of measurement in quantum mechanics is proposed. In this approach, the process of measurement is described in the Heisenberg picture and divided into two stages. The first stage is the transduction from the measured observable to the probe observable. The second stage is the amplification of the probe observable to the macroscopic meter observable. Transduction in the Heisenberg picture is the counterpart of entanglement in the Schrödinger picture. The object-apparatus interaction is turned on during the first stage and described by the unitary time evolution of the composite system. Quantum state reduction is thus derived from this interaction within quantum mechanics based on the quantum Bayes principle — the Bayes principle applied to quantum mechanical joint probability and determining quantum state [7]. The dynamical process of the second stage is described as a quantum amplification with *infinite gain*. A mathematically precise description of a quantum amplification with infinite gain is given by nonstandard extension of quantum mechanics [8], in which classical mechanics and quantum mechanics are synthesized in terms of mathematical device of nonstandard analysis.

For simplicity, we will be confined to measurements of *discrete observables*, but it will be easy for the reader to generalize the argument to continuous observables and to join the argument to the general theory developed in such papers as [3, 4, 9, 10, 11].

## 2 New approach to describing measuring processes

Let **S** be the measured object, and **A** an apparatus measuring an observable $A$ with purely discrete spectrum of the object **S**. The Hilbert spaces of **S** and **A** are denoted by $\mathcal{H}_\mathbf{S}$ and $\mathcal{H}_\mathbf{A}$, respectively. The process of measurement of $A$ at a time $t$ using the apparatus **A** is described as follows. The interaction between **S** and **A**, called the *measuring interaction*, is turned on from the time $t$ to a later time $t + \Delta t$ and that after $t + \Delta t$ the object is free from the apparatus. The outcome of measurement is obtained by measuring the probe observable $B$ in the apparatus at the time $t + \Delta t$, that is described as the time evolution of the apparatus from the time $t + \Delta t$ to a later time $t + \Delta t + \tau$; hence, it is a local measurement of $B$ at the system **A** [7]. At the time $t + \Delta t + \tau$ the observer reads out directly the value of the macroscopic observable $C$ in the apparatus **A** as the outcome of the measurement of $A$ at the time $t$ using the apparatus **A**. The time $t$ is called the *time of measurement*, the time $t + \Delta t$ is called the *time just after measurement*, and the time $t + \Delta t + \tau$ is called the *time of readout*.

Thus, the process of measurement is divided into two stages. The first stage, from the time $t$ to the time $t + \Delta t$, transduces the measured observable to the probe observable. The second stage, from the time $t + \Delta t$ to the time $t + \Delta t + \tau$, amplifies the probe observable to the macroscopic meter observable. Both stages are characterized well in the Heisenberg picture as follows. Denote by $E^A(a)$ the projection operator of the Hilbert space $\mathcal{H}_\mathbf{S}$ with the range $\{\psi \in \mathcal{H}_\mathbf{S}|\ A\psi = a\psi\}$. Suppose that the time evolution of the composite system $\mathbf{S} + \mathbf{A}$ from the time $t$ to the time $t + \Delta t$ is represented by a unitary operator $U$ on the Hilbert space $\mathcal{H}_\mathbf{S} \otimes \mathcal{H}_\mathbf{A}$. Suppose that at the time of measurement the object **S** is in an arbitrary state $\psi$ and that the apparatus **A** is in a fixed state $\xi$. We describe the time evolution of the composite system $\mathbf{S} + \mathbf{A}$ in the Heisenberg picture taking the time $t$ as the original time with the original state $\psi \otimes \xi$. The Heisenberg operators $A(t)$, $B(t+\Delta t)$ are, thus, given by $A(t) = A \otimes I$ and $B(t+\Delta t) = U^\dagger(I \otimes B)U$.

The condition for the first stage of the the above model to describe a measurement of $A$ is as follows: *The measured observable $A$ and the probe observable $B$ are related by the measuring interaction as*

$$B(t+\Delta t)(\psi \otimes \xi) = A(t)(\psi \otimes \xi) \tag{1}$$

*for arbitrary $\psi$.* If the above condition is satisfied, we say that *the observable $A$ of $\mathbf{S}$ is transduced to the observable $B$ of $\mathbf{A}$ by the time evolution $U$ with the preparation $\xi$ of $\mathbf{A}$*. In this case, the time evolution from $t$ to $t+\Delta t$, or the unitary operator $U$, is called the *transduction* of $A$ to $B$ with $\xi$, and we call the above condition, (1), as the *transduction requirement*. Therefore, we require precisely that the first stage of measurement is the transduction from the measured observable to the probe observable with the fixed apparatus preparation.

If we introduce the *noise operator $N$* defined by $N = B(t+\Delta t) - A(t)$, the transduction requirement is equivalent to the condition

$$N(\psi \otimes \xi) = 0 \tag{2}$$

for arbitrary $\psi$. It is easy to see that the transduction requirement, (1), is equivalent to the requirement

$$\Pr\{A(t) = a\} = \Pr\{B(t+\Delta t) = a\} \tag{3}$$

for arbitrary object state at the time $t$. From the Born statistical formula, (3) is equivalent to

$$\langle \psi | E^A(a) | \psi \rangle = \langle \psi \otimes \xi | U^\dagger (1 \otimes E^B(a)) U | \psi \otimes \xi \rangle. \tag{4}$$

Since $\psi$ is arbitrary, it is concluded that (1) is equivalent to the relation

$$E^A(a) = \mathrm{Tr}_{\mathbf{A}}[(1 \otimes E^B(a)) U (I \otimes |\xi\rangle\langle\xi|) U^\dagger], \tag{5}$$

where $\mathrm{Tr}_{\mathbf{A}}$ denotes the partial trace over $\mathcal{H}_{\mathbf{A}}$. The above relation is the sole requirement proposed previously in [3] for $(\mathcal{H}_{\mathbf{A}}, \xi, U, B)$ to describe a measuring process of $A$

A (sufficient) condition, to be proposed in this paper, for the second stage of the present model to describe a measurement of $A$ is as follows: *There is a positive infinite c-number $G$ such that*

$$C = G^{-1} B \tag{6}$$

*and that*

$$B(t+\Delta t + \tau)(\psi \otimes \xi) = GB(t+\Delta t)(\psi \otimes \xi) \tag{7}$$

*for arbitrary $\psi$.*

The first condition, (6), implies that the observable $C$ is macroscopic and it will be referred to as the *macro-meter requirement*. To see this, suppose that $B$ has the canonically conjugate observable $B'$ such that

$$[B, B'] = -i\hbar I. \tag{8}$$

Let $u_q$ be the microscopic unit of the quantity represented by $B$ and let $u_c$ be the corresponding macroscopic unit such that $u_c = Gu_q$. Since $G$ is positive infinite, we have $u_c \gg u_q$. Then, we have $Cu_c = Bu_q$ so that the observable $C$ can be considered to represent the same physical quantity as $B$ in the macroscopic unit. Similarly, let $u'_q$ be the microscopic unit for the quantity represented by $B'$ and let $u'_c$ be the corresponding macroscopic unit such that $u'_c = G'u'_q$, where $G'$ is supposed to be an positive infinite c-number. Letting $C' = G'^{-1}B'$, we have $C'u'_c = B'u'_q$ so that the observable $C'$ can be considered to represent the same physical quantity as $B'$ in the macroscopic unit. Then, by (8) we have

$$[C, C'] = -i\hbar G^{-1} G'^{-1} I. \tag{9}$$

Since $\hbar$, the Planck constant in the microscopic unit divided by $2\pi$, is a finite c-number and since $G^{-1}$ and $G'^{-1}$ are an infinitesimal c-number, the commutator $[C, C']$ is an infinitesimal c-number, i.e., $[C, C'] \approx 0$, so that the operators $C$ and $C'$ commute each other by neglecting infinitesimal numbers. Since every macroscopic observable can be regarded as a function of the pair of canonically conjugate observables $C$ and $C'$, all the macroscopic observables commute each other. Thus, it is reasonable to consider the observable $C$ to be macroscopic.

The second condition, (7), requires that the observable $B$ is amplified by the time evolution of the apparatus from the time just after measurement to the time of readout and it will be referred to as the *infinite amplification requirement*.

From (6) and (7), we have

$$C(t + \Delta t + \tau)(\psi \otimes \xi) = B(t + \Delta t)(\psi \otimes \xi) \tag{10}$$

for arbitrary $\psi$. Hence, the second stage amplifies the microscopic probe observable to the macroscopic meter observable. The requirement for the second stage represented by (6) and (7) is called as the *macroscopic amplification requirement*.

By combining (1) and (10), we have

$$C(t + \Delta t + \tau)(\psi \otimes \xi) = A(t)(\psi \otimes \xi) \tag{11}$$

for arbitrary $\psi$, or equivalently

$$\Pr\{C(t + \Delta t + \tau) = a\} = \Pr\{A(t) = a\} \tag{12}$$

for arbitrary object state at the time $t$. Thus, the measurement of $A$ at the time of measurement is reduced to the reading of the macroscopic meter observable $C$ at the time of readout. In particular, the quantum mechanical probability distribution of the measured observable presupposed by the Born statistical formula in the right-hand side of (12) is construed as the probability distribution of the macroscopic meter observable in the left-hand side of (12).

Theory of infinite and infinitesimal numbers is developed rigorously in nonstandard analysis invented by Robinson [12]. For basic methods of nonstandard analysis and its applications to physics, we refer the reader to [13]. For the precise description of the nonstandard extension of quantum mechanics and its application to the phase operator problem, we refer the reader to [8]. The role of quantum amplifiers in measurements of optical systems was discussed by Glauber [14] previously.

# 3  Quantum state reduction

In this section, it will be shown that the state reduction is determined by the first stage of measurement, specified by the preparation $\xi$ of the apparatus, the unitary $U$ of the measuring interaction, and the probe observable $B$.

In what follows, we consider the time evolution of the state in the Schrödinger picture, and the state is represented by density operators unless stated otherwise. Let $\rho(t) = |\psi\rangle\langle\psi|$, the state of **S** at the time of measurement. At the time just after measurement, the system $\mathbf{S} + \mathbf{A}$ is in the state $U|\psi \otimes \xi\rangle\langle\psi \otimes \xi|U^\dagger$, and hence the object **S** is in the state

$$\rho(t + \Delta t) = \operatorname{Tr}_{\mathbf{A}}[U|\psi \otimes \xi\rangle\langle\psi \otimes \xi|U^\dagger]. \tag{13}$$

Let $\rho(t + \Delta t | A(t) = a)$ be the state at the time just after measurement of the object that leads to the outcome $a$. If $\Pr\{A(t) = a\} = 0$, the state $\rho(t + \Delta t | A(t) = a)$ is not definite, and we let $\rho(t + \Delta t | A(t) = a)$ be an arbitrarily chosen density operator for mathematical convenience.

In order to characterize the state $\rho(t + \Delta t | A(t) = a)$, suppose that at the time $t + \Delta t + t'$ with $t' \geq 0$ the observer were to measure an arbitrary observable $X$ of the same object using an arbitrary apparatus **X** measuring $X$. If $\Pr\{A(t) = a\} \neq 0$, let $\Pr\{X(t + \Delta t + t') = x | A(t) = a\}$ be the conditional probability that the outcome of the $X$-measurement at the time $t + \Delta t + t'$ is $x$ given that the outcome of the $A$-measurement at $t$ is $a$. Then, by the quantum Bayes principle [7] we have

$$\Pr\{X(t + \Delta t + t') = x | A(t) = a\} = \operatorname{Tr}[e^{iHt'/\hbar}E^X(x)e^{-iHt'/\hbar}\rho(t + \Delta t | A(t) = a)], \tag{14}$$

where $H$ is the Hamiltonian of the object **S**. Since $X$ is arbitrary, the density operator $\rho(t + \Delta t | A(t) = a)$ satisfying (14) is at most uniquely determined so that we can regard (14) as the mathematical characterization of the state $\rho(t + \Delta t | A(t) = a)$.

Since the outcome $B(t + \Delta t) = a$ is interpreted as the outcome $A(t) = a$, we have

$$\Pr\{A(t) = a, X(t + \Delta t + t') = x\} = \Pr\{B(t + \Delta t) = a, X(t + \Delta t + t') = x\}. \tag{15}$$

Since the measurement of $B$ at $t + \Delta t$ is local at **A**, by the local measurement theorem we have

$$\Pr\{B(t + \Delta t) = a, X(t + \Delta t + t') = x\}$$
$$= \langle\psi \otimes \xi|U(e^{iHt'/\hbar}E^X(x)e^{-iHt'/\hbar} \otimes E^B(a))U^\dagger|\psi \otimes \xi\rangle. \tag{16}$$

Thus, the conditional probability of the outcome $X(t + \Delta t + t') = x$ given $A(t) = a$ satisfies

$$\Pr\{X(t + \Delta t + t') = x | A(t) = a\}$$
$$= \frac{\Pr\{B(t + \Delta t) = a, X(t + \Delta t + t') = x\}}{\sum_a \Pr\{B(t + \Delta t) = a, X(t + \Delta t + t') = x\}}$$
$$= \frac{\langle\psi \otimes \xi|U(e^{iHt'/\hbar}E^X(x)e^{-iHt'/\hbar} \otimes E^B(a))U^\dagger|\psi \otimes \xi\rangle}{\langle\psi \otimes \xi|U(I \otimes E^B(a))U^\dagger|\psi \otimes \xi\rangle}. \tag{17}$$

Since $X$ is arbitrary, from (14) we have

$$\rho(t + \Delta t | A(t) = a) = \frac{\operatorname{Tr}_{\mathbf{A}}[(I \otimes E^B(a))U|\psi \otimes \xi\rangle\langle\psi \otimes \xi|U^\dagger]}{\operatorname{Tr}[(I \otimes E^B(a))U|\psi \otimes \xi\rangle\langle\psi \otimes \xi|U^\dagger]}. \tag{18}$$

The defining equation (14) shows that the posterior distribution of the outcome of the measurement of *any* observable $X$ of the object **S** after the time $t+\Delta t$ is described by the state $\rho(t+\Delta t|A(t)=a)$. Therefore, we can conclude that the information $A(t)=a$ changes the state of the system **S** from the *prior state* $\rho(t+\Delta t)$ to the *posterior state* $\rho(t+\Delta t|A(t)=a)$ according to the quantum Bayes principle.